\documentclass[twocolumn,showpacs,preprintnumbers,amsmath,amssymb]{revtex4}
\usepackage{graphicx}
\usepackage{epstopdf}
\usepackage{amssymb}
\usepackage{MnSymbol}
\usepackage{dcolumn}
\usepackage{bm}
\usepackage{dsfont}

\begin{document}


\title{Systems poised to criticality through Pareto selective forces}

\author{Lu\'is F. Seoane$^{1,2,3}$ and Ricard Sol\'e$^{2,3,4}$ }
\affiliation{
  $^1$Department of Physics, Massachusetts Institute of Technology, 77 Massachusetts Ave, Cambridge, MA 02139, USA. 
  \\$^2$ICREA-Complex  Systems   Lab,  Universitat Pompeu Fabra  -  PRBB, Dr. Aiguader 88, 08003  Barcelona, Spain. 
  \\$^3$Institut de Biologia Evolutiva, UPF-CSIC, Barcelona. 
  \\$^4$Santa Fe Institute, 1399 Hyde Park Road, New Mexico 87501, USA. 
  }

\begin{abstract}

  Pareto selective forces optimize several targets at the same time, instead of single fitness functions. Systems subjected to these forces evolve towards their Pareto front, a geometrical object akin to the thermodynamical Gibbs surface and whose shape and differential geometry underlie the existence of phase transitions. In this paper we outline the connection between the Pareto front and criticality and critical phase transitions. It is shown how, under definite circumstances, Pareto selective forces drive a system towards a critical ensemble that separates the two phases of a first order phase transition. Different mechanisms implementing such Pareto selective dynamics are revised.

\end{abstract}


\maketitle

  Critical systems are characterized by physical quantities that diverge as $\theta \sim 1 / |T-T_c|^\delta$ when $T \rightarrow T_c$. Criticality often requires a careful handling of control parameters, e.g. percolation probability must be set to $p = p_c$ or water becomes opalescent only at pressure and temperature $(P_c,\> T_c)$. Despite this fine tuning problem, evidence for criticality is common in complex systems, including written texts \cite{Zipf1949}, populations in cities \cite{Auerbach1913} or cascading events \cite{LuHamilton1991, BeggsPlenz2003, LippielloArcangelis2012, HaimoviciChialvo2013}. While the hypothesis that some complex systems might be poised to criticality \cite{Kauffman1993, MoraBialek2011} has been controversial, it is supported by several existing mechanisms known to induce power-laws in a robust manner \cite{BakWiesenfeld1987, BakWiesenfeld1988, Bak1996, CarlsonDoyle1999, CarlsonDoyle2000, Mitzenmacher2004, Newman2005}. These power laws are often informally linked to potential evolutionary paths leading to optimality, but a deep connection between criticality and evolved structures is still largely missing. In this letter we aim to provide one potential avenue to such connection.

  A first piece for our approach is the route to criticality provided by {\em Maximum Entropy} (MaxEnt) models. Take some measurements ($\{f_k\}, \> k = 1, \dots, K$) performed on a system that can be found in any of ${\boldsymbol \sigma}_j \in \Sigma$ microstates. Among all possible probability distributions $\{P_i({\boldsymbol \sigma}_j)\}$, the one that most faithfully describes our observations $\{f_k\}$ is the one with largest entropy and takes the form \cite{Jaynes1957a, Jaynes1957b}:
    \begin{eqnarray}
      P\left({\boldsymbol \sigma}_j, \{\lambda_k\}\right) &=& {1 \over Z} \exp\left({-\sum_k \lambda_k f_k \left({\boldsymbol \sigma}_j \right)}\right), 
      \label{eq:1}
    \end{eqnarray}
  where $\lambda_k$ are inferred from the data and $Z = \sum_j \exp({-\sum_k \lambda_k f_k({\boldsymbol \sigma}_j)})$. MaxEnt has been successfully applied to diverse complex systems: letters within words \cite{StephensVialek2010}, antibody coding regions \cite{MoraCallan2010}, pixels in natural images \cite{StephensBialek2013}, or spiking neurons  \cite{TkacikBialek2013, TkacikBerry2014, TkacikBialek2015}. From (\ref{eq:1}) it is possible to evaluate susceptibilities $\chi_{kl} = \partial \left< f_k \right>_P/\partial \lambda_l$, where $\left< f_k \right>_P$ is the average value of the observable $f_k$ given the model $P\left({\boldsymbol \sigma}_j, \{\lambda_k\}\right)$. These $\chi_{kl}$ measure responses of a system to changes of the control parameter $\lambda_l$. Diverging $\chi_{kl}$ are at the core of the power-law behaviors at criticality.

   \begin{figure}
      \begin{center}
        \includegraphics[width = \columnwidth]{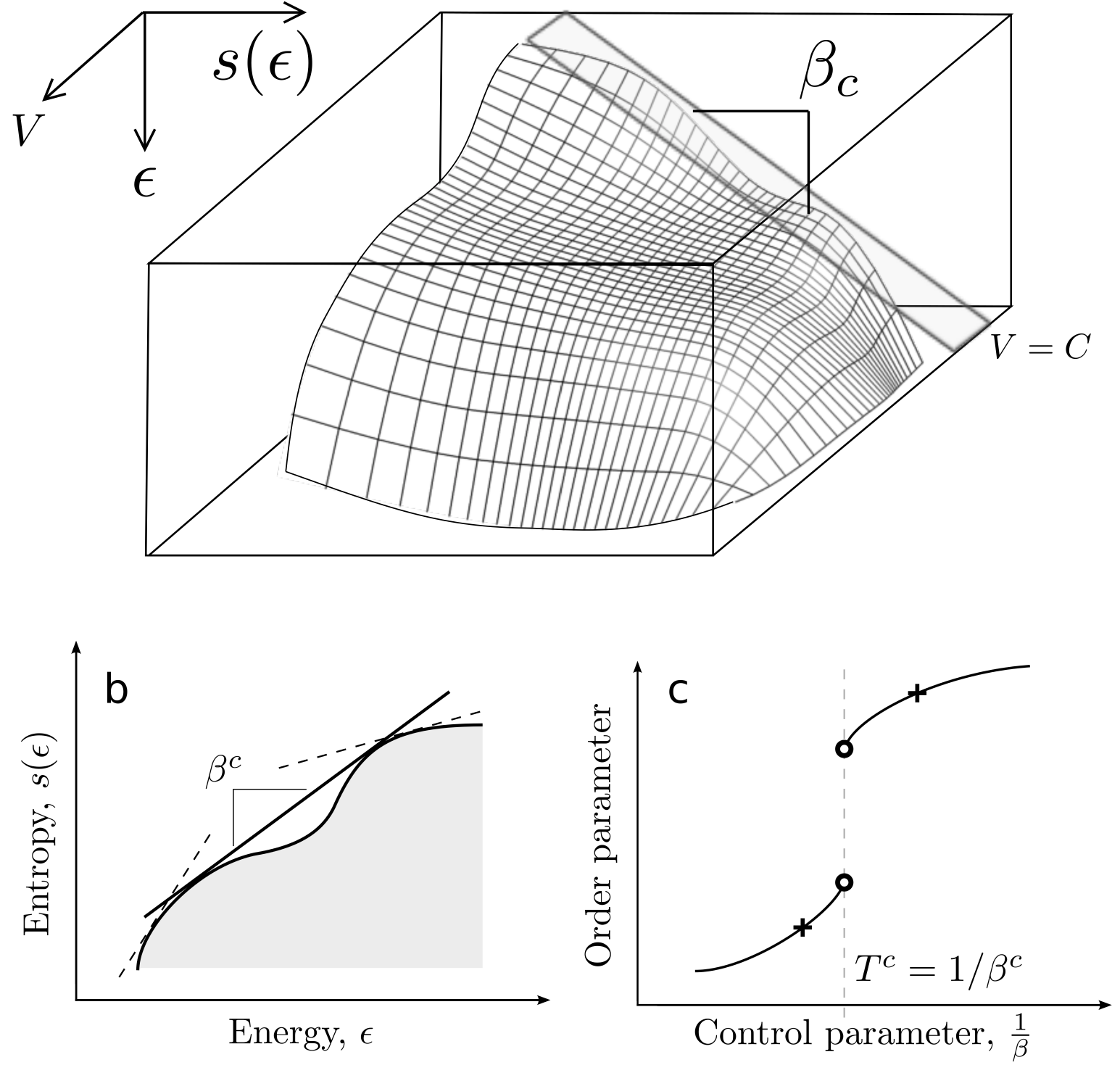}

        \caption{\textbf{Phase transitions in the Gibbs surface. } {\bf a} Microcanonical ensembles of a thermodynamic system conform a $2$-D surface with cavities and edges associated to phase transitions. Canonical ensembles constitute the convex hull, obtained by rolling a rigid plane over this surface. {\bf b} If $V$ is irrelevant or constant, a cross section (energy-entropy plot) accounts for the thermodynamics of the system. $\beta = 1/T$ is the slope of the tangent line. {\bf c} States away from a first order transition (crosses, found at the tangents of the dashed lines in panel {\bf b}) and coexisting phases (circles, found at the two tangent points of the solid line in panel {\bf b}) are separated by a gap in any order parameter. }

        \label{fig:1}
      \end{center}
    \end{figure}

  Secondly, we use the Gibbs surface (Fig. \ref{fig:1}{\bf a}) \cite{Gibbs1873b, Maxwell1904}. Each point of this object is linked to a microcanonical ensemble description of the system. Macroscopic equilibrium states lie strictly on the convex hull of that surface, whose coordinates correspond to $(P, T, V)$ thermodynamic states. Hence solving equilibrium thermodynamics is equivalent to rolling a rigid plane over the Gibbs surface (Fig. \ref{fig:1}{a,b}). The slope of this plane reads $T$ and $P$ at each equilibrium point. If the surface presents a cavity, the system bypasses it at some value of the control parameters (Fig. \ref{fig:1}{\bf b,c}) leading to a first order phase transition. Second order transitions take place if the Gibbs surface presents sharp edges \cite{SeoaneSole2013}.
      
\begin{figure*}
    \begin{center}
      \includegraphics[width = \textwidth]{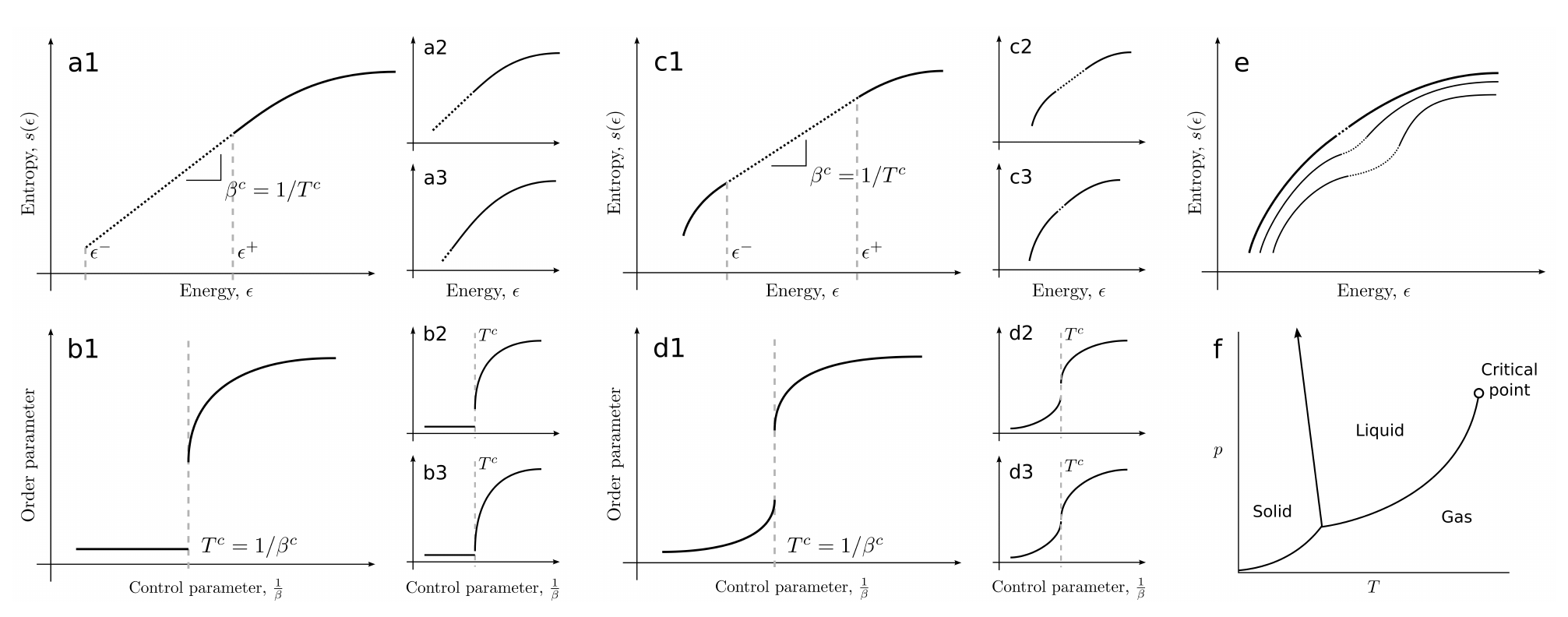}

      \caption{\textbf{A geometric overview of criticality.} {\bf a-d} Systems critical over an energy range $[\epsilon^-, \epsilon^+]$ have linear entropies as a function of $\epsilon$. {\bf 1-3} Their limits as $\epsilon^- \rightarrow \epsilon^+$ yield well known critical scenarios. {\bf b}, {\bf d} Plots of order parameters reveal {\em hybrid phase transitions} with features of both first and critical transitions \cite{DorogovtsevMendes2006, BaxterMendes2010, HuHavlin2011, WuHu2014}. The nature of each of these cases can be clarified when looking at the Pareto front. {\bf e} Another path to the critical point in {\bf c1-3} is possible in $2$-D Gibbs surfaces and accounts for the classic critical point of the liquid-vapor transition {\bf f}. }

      \label{fig:2}
    \end{center}
  \end{figure*}

  When volume is not relevant the Gibbs surface is a curve relating entropy and energy [$s = \bar{s}(\epsilon)$] and the rolling plane is a straight line with slope $\beta = 1/T$ (Fig. \ref{fig:2}{\bf b,c}). The only relevant susceptibility then is heat capacity:
    \begin{equation}
      C = {N \over T^2}\left( -{d^2 \bar{s}(\epsilon) \over d\epsilon^2} \right)^{-1}. 
      \label{eq:2}
    \end{equation}
  Criticality translates into a geometric condition for $\bar{s}(\epsilon)$ \cite{MoraBialek2011, TkacikBialek2013, TkacikBialek2015}, namely $C \rightarrow \infty$ if $d^2 \bar{s}(\epsilon)/ d\epsilon^2 = 0$. If this condition holds for every energy, then $\bar{s}(\epsilon) = A + B\epsilon$ is a straight line. Such extreme critical behavior has been reported as the thermodynamic limit of empirical models of certain spiking neurons \cite{TkacikBialek2013, TkacikBialek2015}. Less radically, $d^2 \bar{s}(\epsilon) / d\epsilon^2 = 0$ for an energy range $\epsilon \in [\epsilon^-, \epsilon^+]$ (Fig. \ref{fig:2}{\bf a1}, {\bf c1}) is seemingly found in MaxEnt models of natural images (Fig. 3 in \cite{StephensBialek2013}). When rolling a rigid line over such Gibbs surfaces the critical range $[\epsilon^-, \epsilon^+]$ also implies a drastic rearrangement of the system from energy $\epsilon^-$ to $\epsilon^+$ at the critical temperature, resulting in a gap in any order parameter (Figs. \ref{fig:2}{\bf b1}, {\bf d1}). This behavior contains elements of both critical and first order transitions, which is known as {\em hybrid} phase transition \cite{DorogovtsevMendes2006, BaxterMendes2010, HuHavlin2011, WuHu2014}. Shrinking the relevant range  ($\epsilon^- \rightarrow \epsilon^+$) results either in a critical second order transition (Fig. \ref{fig:2}{\bf a1-3}, {\bf b1-3}) or just one critical point (Figs. \ref{fig:2}{\bf c1-3} and {\bf d1-3}). The latter case also happens if a cavity vanishes, as seen in $2$-D Gibbs surfaces for many liquid-gas transitions that cease to exist (Figs. \ref{fig:2}{\bf e,f}).
      
\begin{figure}
    \begin{center}
      \includegraphics[width = 6.5 cm]{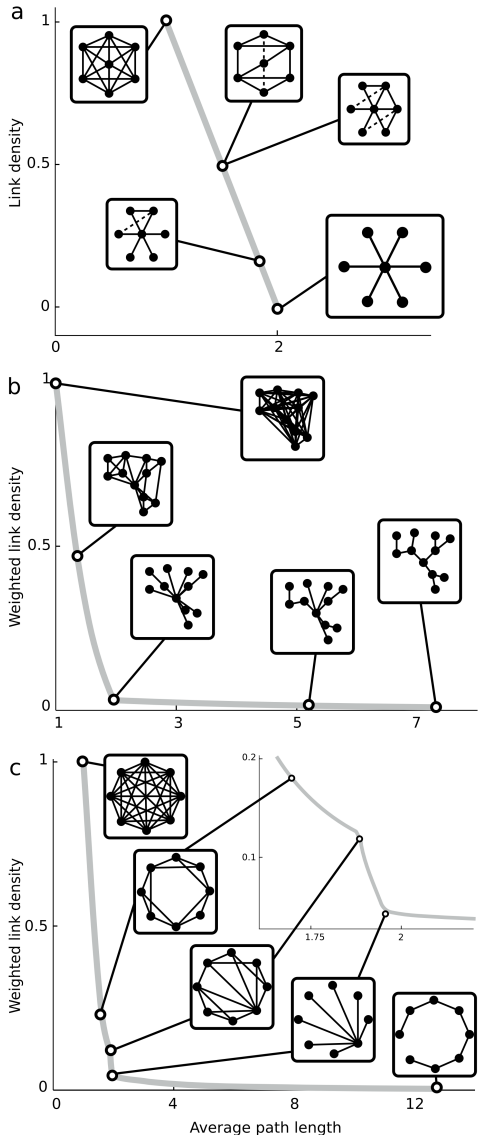}

      \caption{\textbf{Simultaneous minimization of average path length and edge density.} Graph drawings are qualitative but the Pareto fronts are faithful. The scales of the axes are explained in \cite{SeoaneSole2015a}, but do not affect the current discussion. {\bf a} Purely topological graphs lie on linear Pareto front. In {\bf b}, {\bf c} the Euclidean distance weights the cost of each link when computing the density of edges. {\bf b} Nodes distributed over a plane. Optima trade between a clique, a star graph, and the Minimum Spaning Tree through two second order transitions. {\bf c} Nodes placed over a circle display a first and a second order transition. }

      \label{fig:3}
    \end{center}
  \end{figure}

  Now we use the third component of our analysis: the so called Pareto front, which provides a geometrical characterization of optimal solutions and is connected to  thermodynamics \cite{SeoaneSole2013, SeoaneSole2015a, SeoaneSole2015b, Seoane2016}.  Pareto Optimality (PO) deals with the simultaneous minimization of a set of {\em target functions} $T_f \equiv \{t_k, \> k=1, \dots, K\}$ which often conflict with each other leading to non-trivial optima. Consider e. g. the set $\Gamma$ of all connected networks with $N$ nodes. We seek to find the subset $\Pi_\Gamma \subset \Gamma$ of graphs that simultaneously minimize the targets: i) average path length ($t_1$) and ii) density of edges ($t_2$). Given two networks $\gamma_1, \gamma_2 \in \Gamma$, we say that $\gamma_1$ dominates $\gamma_2$ ($\gamma_1 \prec \gamma_2$) if:
    \begin{eqnarray}
      t_k(\gamma_x) &\le& t_k(\gamma_y) \>\> \forall \>\> k=\{1,2\}; \nonumber \\ 
      \exists \> k'\in\{1,2\} &|& t_{k'}(\gamma_x) < t_{k'}(\gamma_y). 
      \label{eq:3}
    \end{eqnarray}
  If $\gamma_1 \prec \gamma_2$, then $\gamma_1$ is objectively better than $\gamma_2$ regarding our targets. Often $\gamma_1$ and $\gamma_2$ are mutually non-dominated and we cannot choose between them -- unless we introduce an unjustified bias, which we wish to avoid by now. A network not dominated by any other is Pareto optimal. The set $\Pi_\Gamma \subset \Gamma$ consists of all Pareto optimal networks (hence all $\gamma \in \Pi_\Gamma$ are mutually non-dominated). Plotting $(t_1, t_2)$ for every $\gamma \in \Pi_\Gamma$ renders the most optimal tradeoff between targets (Fig. \ref{fig:3}), called the {\em Pareto front}. (See methods and technical details elsewhere \cite{Coello2006, FonsecaFleming1995, Dittes1996, Zitzler1999, KonakSmith2006}.) If we build the simplest possible {\em global energy} with our targets ($\Omega \equiv \sum_k \lambda_k t_k$), phase transitions arise (as with the Gibbs surface) due to the cavities and edges of $\Pi_\Gamma$ \cite{SeoaneSole2013, SeoaneSole2015a, SeoaneSole2015b, Seoane2016}. We propose now to apply also the geometric  condition for criticality to Pareto optimal systems.

  In \cite{SeoaneSole2015a} we studied variations upon the Pareto optimal complex networks introduced above. Two targets are present so
    \begin{eqnarray}
      \min\left\{\sum_k t_k \lambda_k \right\} &\equiv& \min \left\{ \Omega = \lambda t_1(\gamma) + (1-\lambda)t_2(\gamma) \right\}. 
      \label{eq:4}
    \end{eqnarray}
  Here we consider three cases: {\bf A},  minimization of both topological path length and edge density  (geometric space is irrelevant). In this case $\Pi_{\Gamma}$ is a straight line with  slope $d = -1$ (Fig. \ref{fig:3}{\bf a}), which implies a first order phase transition with a critical point (following the criterion above) at $\lambda = \lambda^c_A \equiv -d/(1-d) = 1/2$. For cases {\bf B} and {\bf C}, each link still contributes the same to the average path length, but edge density is weighted by the Euclidean distance.  In {\bf B} nodes are scattered randomly over $[0,1] \times [0,1] \in \mathds{R}^2$. Here $\Pi_{\Gamma}$ (Fig. \ref{fig:3}{\bf b}) presents two sharp edges (see \cite{SeoaneSole2015a}), implying two second order transitions. In {\bf C}, nodes are regularly spaced over a circle. The front (Fig. \ref{fig:3}{\bf c}) presents a cavity (thus a first order transition) and a sharp edge (hence a second order transition -- again, see \cite{SeoaneSole2015a} for details).

\begin{figure}
    \begin{center}
      \includegraphics[width = \columnwidth]{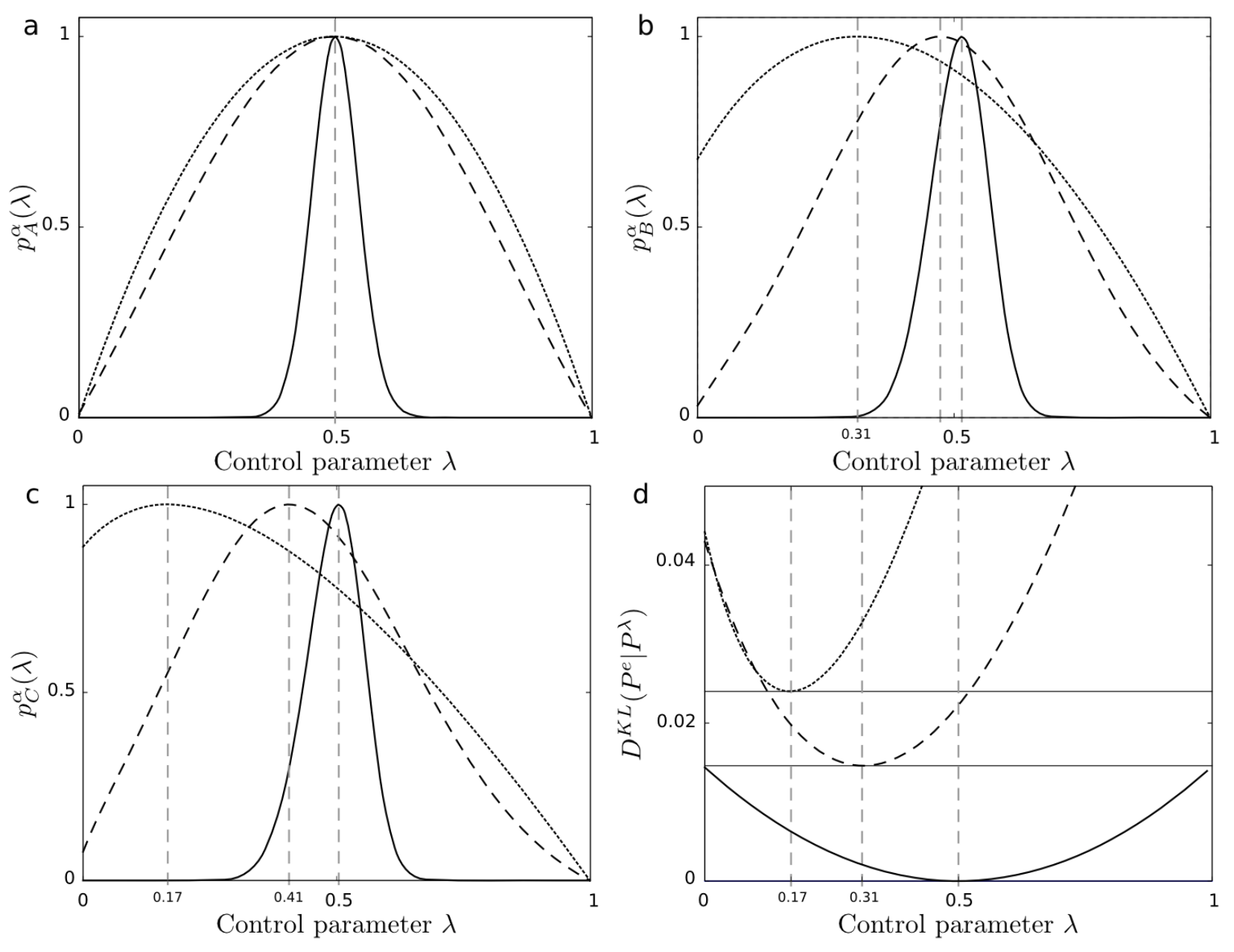}

      \caption{\textbf{Testing MaxEnt models. } {\bf a-c} $p$-values for MaxEnt models of Pareto optimal sets $\Pi_{A,B,C}$ as a function of $\lambda$ and $\alpha$ ($\alpha = 0.0125$, solid; $\alpha = 0.1$, dashed; and $\alpha = 1$ dotted lines; curves have been normalized for comparison). {\bf a} $\alpha$ does not affect the critical ($\tilde{\lambda}_A = \lambda^c_A$) description of $\Pi_A$. {\bf b}, {\bf c} Changing $\alpha$ changes the best model, so that an $\alpha$-invariant, consistent description does not arise for these Pareto optima. $\tilde{\lambda}_{B,C}(\alpha)$ usually do not correspond to relevant parameters in phase space. {\bf d} The best model misses the least information about each data set ($\Pi_A$, solid; $\Pi_B$, dashed; and $\Pi_C$ dotted lines; $\alpha=1$). This loss is vanishingly small in the critical case.}

      \label{fig:4}
    \end{center}
  \end{figure}

  As defined, $\Pi_{\Gamma}$ constitutes a set that we can study statistically as a thermodynamical ensemble -- similarly to how other empirical data sets have been studied \cite{StephensVialek2010, MoraCallan2010, StephensBialek2013, TkacikBialek2013, TkacikBerry2014, TkacikBialek2015}. A MaxEnt model of the kind:
    \begin{eqnarray}
      P^\lambda(\gamma; \alpha) &=& {1 \over Z} \exp(-\alpha\Omega(\gamma, \lambda)), 
      \label{eq:5}
    \end{eqnarray}
  with $\alpha$ an arbitrary scaling factor, estimates how often we find each network in the ensemble. We compute
    \begin{eqnarray}
      \tilde{\lambda}(\alpha) &=& \max_\lambda \left\{ p^\alpha_{A,B,C}(\lambda) = \prod_{\gamma \in \Pi_\Gamma} \exp(-\alpha\Omega(\gamma, \lambda)) \right\}
      \label{eq:6}
    \end{eqnarray}
  to obtain the most likely model given the data. Fig. \ref{fig:4}{\bf a-c} shows $p$-values $p^\alpha_{A,B,C}(\lambda)$ for different $\alpha$. These $p$-values (introduced in the argument of the maximization problem in equation \ref{eq:6}) have a large value if the corresponding MaxEnt model renders a good description of $\Pi_\Gamma$. Note how only $\tilde{\lambda}_A$ remains unchanged at the critical value of the system ($\tilde{\lambda}_A = \lambda^c_A$), suggesting that $\Pi_A$ prefers that singular statistical description under every circumstance. Meanwhile, $\tilde{\lambda}_{B,C}$ change depending on $\alpha$ and do not correspond to relevant parameters in the phase space.

  Alternatively, we use the Kullback-Leibler divergence to measure the information loss when describing Pareto optimal sets through the model in Eq. \ref{eq:5}:
    \begin{eqnarray}
      D^{KL}(P^e||P^\lambda(\alpha)) &=& \sum_{\gamma \in \Pi_\Gamma} P^e(\gamma)log \left({P^e(\gamma) \over P^\lambda(\gamma; \alpha)}\right), 
      \label{eq:7}
    \end{eqnarray}
  where $P^e_{A,B,C}(\gamma) = 1/||\Pi_{A,B,C}||$ is the empirical frequency of each network in the front. This quantity (Fig. \ref{fig:4}{\bf d}) is minimal but not zero at $\tilde{\lambda}_{B,C}(\alpha)$, while it vanishes at $\tilde{\lambda}_{A}(\alpha) = \lambda^c_A$. Again $\min_\lambda \{D^{KL}_{B,C}\}$ (but not $\min_\lambda \{D^{KL}_{A}\}$) depend on $\alpha$ (not shown).

  It turns out that a MaxEnt model tuned to the critical point is a good description of $\Pi_{\Gamma}$ in case {\bf A}. Put otherwise, the Pareto selective forces have driven the network ensemble to a critical state, as seen from a statistical mechanics viewpoint, and the criticality of this ensemble is robust. Whenever a Pareto front is a straight line, the Pareto optimal set (and, consequently, the final result of a Pareto optimal evolutionary process) will always appear tuned to a critical point. The same critical behavior is found in a model of language evolution \cite{SeoaneSole2015b, FerrerSole2003a, ProkopenkoPolani2010, SalgeProkopenko2013, SoleSeoane2014, Seoane2016, SeoaneSole2017} where hearers and speakers attempt to minimize different costs associated to efficient communication. The conflict arising from the simultaneous minimization of  both hearer and speaker efforts leads to a critical front, consistently with previous studies.  This model is just a theoretical approximation, but arguments from game theory also indicate that human language should not contain synonyms \cite{NowakKrakauer1999}, precisely the optimality condition that defines this Pareto front \cite{ProkopenkoPolani2010, SeoaneSole2015b}.

  Do Pareto selective forces (like the ones simulated in \cite{SeoaneSole2015a, SeoaneSole2015b, Seoane2016}) exist in nature? Note first that constrained optimization can reconstruct a Pareto front: Find the networks $\gamma^1 \in O_1, \> \gamma^2 \in O_2, \> \gamma^3 \in O_3, \> \dots$ ($O_i \subset \Gamma$) with a fixed number of nodes ($N$), with precisely $L_1, L_2, L_3, \dots$ edges respectively (with $L_i \in [N-1, N(N-1)/2]$ randomly distributed integers), and with the lowest average path length possible given each $L_i$. The set $O_1 \cup O_2 \cup O_3 \cup \dots$ samples from the Pareto front in Fig. \ref{fig:3}{\bf a} and will appear critical if described through Eq. \ref{eq:5}. {\em Highly Optimized Tolerance}  ({\em HOT}) states \cite{CarlsonDoyle1999, CarlsonDoyle2000} have been proposed as an alternative to {\em Self-Organized Criticality} \cite{BakWiesenfeld1987, BakWiesenfeld1988, Bak1996} or {\em edge of chaos} dynamics \cite{Kauffman1993} to explain  power-laws in complex systems without resorting to critical states. HOT generates power-laws through thoughtful design that optimizes several constraints simultaneously. One of the HOT strategies is, precisely, constrained optimization \cite{CarlsonDoyle1999, CarlsonDoyle2000}. Disregarding the way to achieve HOT designs, since they solve a PO problem, if the corresponding front is flat these designs will appear critical from the perspective defended in this paper -- which is robustly linked to statistical mechanics. We propose that PO might help close the theoretical gap between SOC and HOT and put under the same light these seemingly confronted approaches. HOT was introduced as a strategy {\em explicitly opposed} to criticality, so we find promising that this case can also be brought under a same theoretical framework when the optimization targets and PO are taken into account carefully. 

  Recent works \cite{NoorMilo2012, ShovalAlon2012, SchuetzSauer2012, HigueraMora2012, SzekelyAlon2013, OteroMurasBanga2014} account for a series of dimension reduction and allometric scalings in real biological data using PO. This suggests that Pareto selective forces might be operating. Numerical evidence also indicates that ecological populations will evolve towards a Pareto front through prey-predator dynamics when different predators select preys with different criteria \cite{LaumannsSchwefel1998, GrimePapaspyrou2012}. If some of these biological systems would belong in a straight Pareto front, based on our results, that system would automatically look critical from a statistical mechanics perspective.

  PO plays a relevant role in economy. The first fundamental theorem of economic welfare guarantees that {\em any competitive market is Pareto efficient at equilibrium} \cite{ArrowDebreu1954, Arrow1963}. If an equilibrium competitive market belongs in a straight Pareto front, it must hence appear critical when studied through MaxEnt methods. The conditions for a {\em competitive market} are stringent and relate to the (often incomplete) available information. However, such markets are an interesting reference of academic importance \cite{Blaug2007, MasGreen1995}.

  Pareto selective forces that consistently poise systems to criticality must relate to power-law distributions, critical exponents, and renormalization invariance. If the optimization targets are energy and entropy and the Pareto front (or the Gibbs surface) is a straight line, a generalized Zipf distribution follows automatically from the most likely MaxEnt model \cite{StephensBialek2013, MoraBialek2011}. Perhaps not surprisingly, Zipf's law is also found (but is not unique) among the Pareto optimal solutions of the least effort language \cite{FerrerSole2003a, ProkopenkoPolani2010, SalgeProkopenko2013, SoleSeoane2014, SeoaneSole2015b, Seoane2016, SeoaneSole2017}. These connections (and those with renormalization) are currently under research for the most general scenarios. 

  Finally, our results should be compared with recent studies concerning empirical evidence of criticality from a data inference perspective. In \cite{MastromatteoMarsili2011} it is shown how complex systems described through MaxEnt models are likely to appear close to a critical point, just because the number of models near this point is larger. Information geometry should be used to correct when estimating such distances to criticality. These issues do not affect our analytical results, which we also extended to two computational models (namely, least effort languages and complex networks -- Fig. \ref{fig:3} and \cite{SeoaneSole2013, SeoaneSole2015a, Seoane2016, SeoaneSole2017}). The robustness of our criticality test puts PO forward as a reliable tool to discuss phase transitions and critical phenomena in complex systems. \\

  We thank the members of the CSL for useful discussions. This work was supported by grants from the Fundaci\'on Bot\'in, the European Research Council (ERC Advanced Grant), and by the Santa Fe Institute.

\end{document}